\documentclass[]{aastex631}

\usepackage{amsmath}

\begin{document}

\title{Investigating the Effects of Atmospheric Stratification on Coronal Active Region Field Modelling}

\author{Oliver Rice (oliver.e.rice@durham.ac.uk)}
\affiliation{Department of Mathematical Sciences, Durham University, Durham, DH1 3LE, UK}

\author{Christopher Prior (christopher.prior@durham.ac.uk)}
\affiliation{Department of Mathematical Sciences, Durham University, Durham, DH1 3LE, UK}

\submitjournal{The Astrophysical Journal}
\accepted{30th Janurary 2025}



\begin{abstract}

Understanding the evolution of the complex magnetic fields found in solar active regions is an active area of research. There are numerous models for such fields which range in their complexity due to the number of known physical effects included in them, the one common factor being they all extrapolate the field up from the photosphere. In this study we focus on the fact that, above the photosphere, and below the corona, lies the relatively cool and dense chromosphere --  which is often neglected in coronal models due to it being comparatively thin and difficult hard to model. We isolate and examine the effect including this boundary layer has on a 2.5D class of driven MHD models of an active region eruption. We find that it can result in significant changes to the dynamics of an erupting field far higher in the atmosphere than the chromosphere itself, generally delaying eruption and increasing the magnetic energy released in each eruption. We also test whether these effects can be approximated using a variation of the more computationally efficient magnetofrictional model, finding a number of simple adaptations of the standard magnetofrictional model capture the effect the chromospheric stratification well. 

\end{abstract}



\section{Introduction} \label{sec:intro}

Active regions in the solar atmosphere are transient areas of particularly strong and complex magnetic field structures. The highly entangled magnetic field in these regions can often release significant amounts of energy, either as solar flares or bursts of plasma in the form of Coronal Mass Ejections (CMEs) \citep{priest2002magnetic}. These events have crucial impacts on space weather \citep{temmer2021space}. The complexity of the magnetic field in active regions is strongly associated with the occurrence of flaring and CME activity and there has been significant interest in modelling such fields (eg. \cite{Wiegelmann2021LRSP,zhu2022magnetohydrostatic,vissers2022active,jiang2016data,leake2017testing,guo2019solar,warnecke2019data,inoue2023comparative,2023NatSR..13.8994T,chen2022comprehensive}).

The scarcity of observed coronal magnetic field data necessitates the use of computational models to reconstruct the solar magnetic field above the photosphere, where measuring the magnetic field is relatively easy. There are a variety of models used to achieve this, varying significantly in complexity. Here we roughly group these into three different classes, increasing in complexity.

The first of these represent the magnetic field as an equilibrium extrapolation from boundary data, which is either observed or simulated but generally assumed to be located on or near the photosphere. The most popular and simplest version of this approach is the `force-free equilibrium', which assumes the magnetic field completely dominates the plasma's behaviour (see \cite{Wiegelmann2021LRSP} for a comprehensive review of these models). Many of these techniques can also be used to find a magnetohydrostatic (MHS) equilibrium (eg. \cite{zhu2018extrapolation,miyoshi2020magnetohydrodynamic,Wiegelmann2021LRSP,zhu2022magnetohydrostatic,vissers2022active}) which takes into account the effect of the plasma pressure in the solar atmosphere. 

The second class of models introduces some time dependence to the plasma's evolution. This can mean solving the full set of magnetohydrodynamic (MHD) equations, subject to magnetic lower boundary conditions specified by observational photospheric data (\cite{jiang2016data,leake2017testing,guo2019solar,warnecke2019data,inoue2023comparative}), or as simulated boundary motions informed by general observations of flows on the surface (\cite{doyle2019observations}). An alternative time-dependent approach is the magnetofrictional model, wherein the fluid equations are replaced with a term which aims to evolve the field towards a force-free state (\cite{guo2016magneto,yardley2018simulating,price2020exploring,2023ApJ...955..114R}). In the absence of dynamic boundary conditions this approach can also be used to recreate static force-free extrapolations.

These modelling approaches require the assumption of a lower boundary fixed at the photosphere, with the boundary condition here not originating self-consistently from the model itself. The final class of models remove this assumption by also including flux emergence through this boundary. These attempt to explain the development of the active regions through by simulating the emergence of the magnetic field (which is often twisted) through the photosphere and up into the solar corona \cite{cheung2014flux,chen2014model,toriumi2019spontaneous,syntelis2019eruptions,2021NatCo..12.6621M,chen2022comprehensive}. The more physically advanced variants include convective motion below and at the photosphere, which can result in significant complexity in the field around the polarity inversion lines. 

There is thus in some sense a hierarchy of magnetic field models, in terms of the physics that they are each able to represent. It can roughly be summarised as follows:
\begin{enumerate}
    \item{The structure of the current density $\textbf{j} =\mu\nabla \times \textbf{B}$. This is present in all models except potential field extrapolations, and is necessary for flaring.}
    \item{The Lorentz force ${\bf j}\times {\bf B}$, which is absent in force-free extrapolations,  but present in all other models. Magnetofrictional models act to minimise the Lorentz Force but it is still present, and is indeed necessary for the evolution of system.}
    \item{Ambient plasma stratification and gravity. This physics is absent from force free extrapolations and magnetofriction. This includes MHD models which make assumptions such as an isothermal plasma.}
    \item{Full plasma dynamics, including convective flow and a non-isothermal atmosphere. This approach can accurately represent the plasma both above and below the transition region in the same model}.
\end{enumerate}
Naturally more complex models than these exist, which (for instance) can include the physics on molecular scales. However, these are of limited use for active region modelling. 

In this paper we focus on the differences between the third and fourth points, namely the effect of properly accounting for a realistic atmospheric stratification in simulations. Whilst MHD simulations necessarily include density and temperature distributions, simplifying assumptions are often made which do not account for the rapid jump in temperature over the chromosphere and transition region  \textit{e.g.} \citep{2009ApJ...691...61P,2013ApJ...778...99L,inoue2018magnetohydrodynamic}. Modelling these effects realistically can be difficult as the transition region occupies a relatively small proportion of the full domain required to model eruptions. It is also very difficult to maintain such stratification self-consistently -- indeed we do not attempt to do so in this study but instead mimic the thermodynamic effects of radiative transfer with a Newton Cooling term. This replaces more sophisticated but computationally very expensive radiative transfer modelling (\textit{e.g.} \cite{toriumi2019spontaneous,2021NatCo..12.6621M}), but has been shown to provide realistic active region magnetic field behaviour \cite{mactaggart2021direct} (\textit{i.e.} it cannot capture radiative emission effectively but provides a realistic field evolution).

In this paper we have two main aims. The first is to investigate the effect of the degree of atmospheric stratification on the evolution of an active region consisting of a magnetic flux rope. In order to isolate and vary systematically the effect of the atmospheric stratification over a significant number of simulations we use a relatively straightforward 2.5D MHD model wherein the flux rope is formed from shearing motions and diffusion on the solar surface (\cite{1989ApJ...343..971V}). We find that over a large number of simulations there is, on average, a significant and increasing delay to the eruptions as the background atmosphere becomes increasingly stratified. In turn this delay leads to a (proportional) increase in the release of magnetic energy with each eruption. 

The second part of this study is to evaluate a series of modifications to the magnetofrictional model and comparing these with the MHD `ground truth'. We find that the addition of an extra fictional `pressure' term to the MF equations can more accurately represent the Lorentz Force distribution of full MHD, and seek to determine which additions to the model can accurately capture the effects of increased stratification: the delay in eruptions and corresponding increase in magnetic energy release. It is hoped these modifications will be of use in developing more realistic global coronal models which require computational efficiency to ensure that whole solar-cycle time periods can be modelled.

\section{MHD Model and Behaviour}

\label{sec:mhd}

We first describe our 2.5D MHD model, which is later used as a `ground truth' against which our magnetofrictional tests can be compared. Our setup is based upon the MHD model used in \cite{2023ApJ...955..114R}, which themselves are based upon the LARE2D code (\cite{2001JCoPh.171..151A}). The simulation domain represents a section of the upper layer of the photosphere, the chromosphere and the lower corona in Cartesian space, within which we observe the formation and eruption of a magnetic flux rope.

Our model requires the assumption of several physical parameters, such as the fluid density, temperature and driving rate at the solar surface. Although we could theoretically attempt to estimate realistic values for these based on observations, there would necessarily exist large uncertainties that would result from this process. Thus as an alternative we instead conduct a reasonably large parameter study, varying both the degree of background stratification and other parameters which independently affect the rate of the formation and eruption of the flux ropes. Due to this need to run many simulations we adopt a relatively simple model which is translationally invariant in the one direction (often known as 2.5D), allowing for far faster computation times than a full 3D model while still exhibiting realistic behaviour.

The domain is chosen to be a square box with dimensions $-0.5 < x < 0.5$, $0 < y < 1$, where the $y$ direction is radial/vertical. The $z$ direction represents the direction of translational symmetry. A distance unit is of the order of magnitude one solar radius ($1 \, R_{\odot}$), and a time unit approximately one day. The majority of the simulation domain lies within the solar corona, with the transition region and chromosphere only at the very base of the domain, below a height of $y^* = 0.0036$. At the resolutions we use in our simulations, this is only on the scale of one or two grid cells.

In dimensionless form (as is the case throughout this paper) the MHD equations used by the LARE2D code are 

\begin{eqnarray}
\frac{\partial \mathbf{B}}{\partial t}&=&-\nabla\times\mathbf{E}\\
\mathbf{E}&=&\eta \mathbf{j} - \mathbf{v}\times\mathbf{B} \\
\mathbf{j}&=& \nabla\times\mathbf{B}\\
\frac{\partial \rho}{\partial t}&=&- \nabla\cdot(\rho \mathbf{v})\\
\frac{\mathrm{D}\mathbf{v}}{\mathrm{D}t}&=&\frac{1}{\rho}\mathbf{j}\times\mathbf{B}
-\frac{1}{\rho}\nabla P + \mathbf{g} \label{eqn:momentum}\\
\frac{\mathrm{D}\epsilon}{\mathrm{D}t}&=&-\frac{P} {\rho}\nabla\cdot\mathbf{v}+\frac
{\eta}{\rho}j^{2} \\
\epsilon &=& \frac{P}{\rho(\gamma-1)} ,
\end{eqnarray}
where the variables are the magnetic field strength $\textbf{B}$, the current density $\textbf{j}$, the fluid (plasma) pressure $P$, the density $\rho$ and the internal energy density $\epsilon$, which is proportional to the temperature of the plasma. The ratio of specific heats is taken to be $\gamma = 5/3$, and we choose a constant gravitational field strength $\mathbf{g} = -1$. Magnetic diffusivity is represented by the constant $\eta = 5 \times 10^{-5}$, chosen to be just high enough to prevent numerical diffusion dominating. We do not use an explicit viscosity, although vicious effects naturally occur to the finite resolution of the code. Our simulations use an evenly-spaced grid at resolution $256 \times 256$.

\begin{figure}[ht!]
    \centering
    \includegraphics[width=\textwidth]{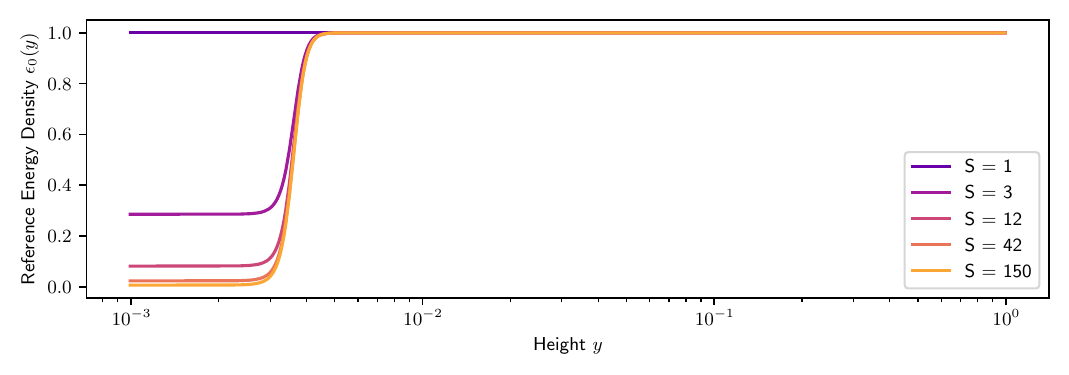}
    \caption{Reference energy density functions $\epsilon_0(y)$, for varying stratification factors $S = 1/T_b$. Note the logarithmic scale on the $x$ axis -- the function is constant throughout almost the entire domain.}
    \label{fig:e0}
\end{figure}

In addition to the standard MHD equations, to  model the atmospheric stratification and the effects this has on the field's evolution require the addition of a dissipative term to the internal energy density equation:
\begin{equation}
\frac{\mathrm{D}\epsilon}{\mathrm{D}t}=-\frac{P} {\rho}\nabla\cdot\mathbf{v}+\frac
{\eta}{\rho}j^{2} + \frac{1}{\rho}\tau(\epsilon_0 - \epsilon), \label{eqn:coolingterm}
\end{equation}
where $\epsilon_0$ is a 1D reference internal energy profile, which will determine the degree of atmospheric stratification. This profile has the form 
\begin{equation}
    \epsilon_0(y) = E_0 \left( 1 + \frac{1}{2}(T_b - 1)(1 - \tanh \left( \frac{y - y^*}{\Delta y} \right) \right),
\end{equation}
where $E_0$ is an overall constant factor, $T_b$ is the temperature of the photosphere relative to the corona, and $\Delta y$ is a parameter which determines the rate of the transition between the two temperatures. This function is plotted in Figure \ref{fig:e0} for a variety of stratification factors $S=1/T_b$. Note that these functions differ only very low in the domain.

This dissipative term encourages the background stratification of the internal energy of the system to return to this state, and prevents the domain from becoming ever cooler as the cool, dense material in the chromosphere is sucked upwards. This approach is similar to that of \cite{2006A&A...450..805L} and \cite{2021GApFD.115...85M}, wherein this simple Newton Cooling term is used to represent the complex non-adiabatic processes which heat the corona, and was found to yield realistic field evolution in \cite{mactaggart2021direct}. We choose the constant parameter $\tau = 0.1$, which is close to the limit at which this cooling term significantly interferes with the magnetic field behaviour. This limit was determined simply by testing a range of values. In the previous work \cite{2023ApJ...955..114R} this extra term was not included, corresponding essentially to $\tau = 0$.

\subsection{Initial Conditions}

As in \cite{2023ApJ...955..114R}, we choose the initial condition for the magnetic field to be a potential (current-free) arcade, with a simple sinusoidal lower boundary condition
\begin{equation}
B_y(x,0) = -B_0 \sin(\pi x),     
\end{equation}
where $B_0 = 1.0$ is a constant (we can still vary the plasma beta in the simulations by varying the initial energy density). We also add a small random fluctuation to remove any numerical symmetry in the system. The initial magnetic field is calculated using a bespoke numerical PFSS  solver, such that the upper boundary condition on the magnetic field is radial.

The initial conditions for the system are determined by first allowing the system to relax into a hydrodynamic equilibrium with no dynamic boundary conditions. The initial internal energy is a constant $E_0$, irrespective of the degree of stratification -- which is imposed solely using the cooling term in Equation \ref{eqn:coolingterm}. 
The initial fluid density $\rho_0$ is then determined by calculating a hydrostatic equilibrium with $\rho_0(y_1) = 1.0$, where $y_1$ is the upper boundary of the domain. 

\begin{figure}[ht!]
    \centering
    \includegraphics[width=0.8\textwidth]{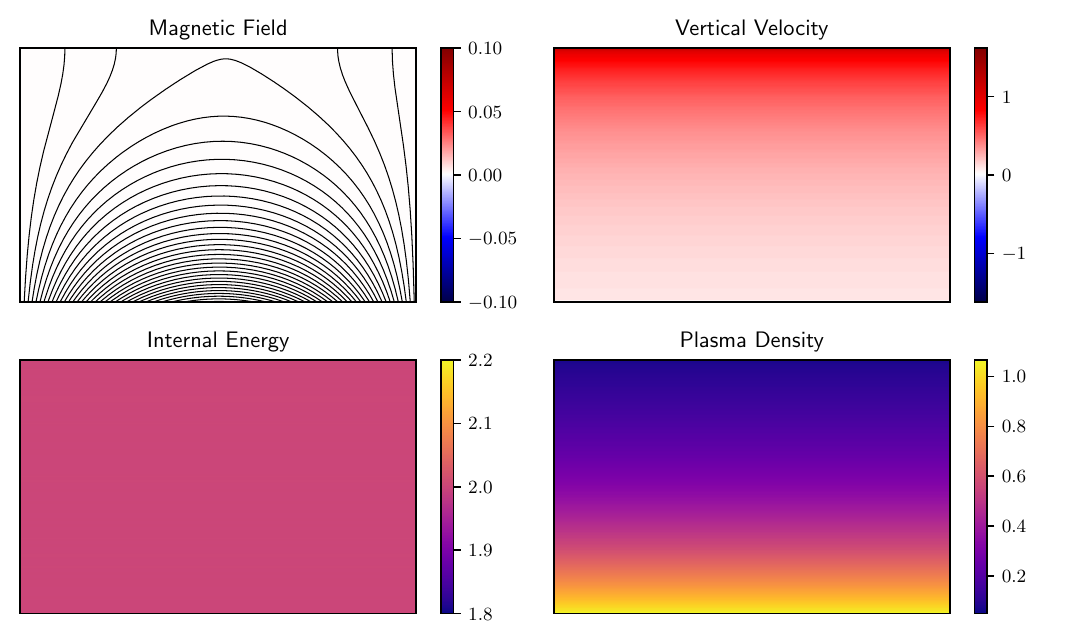}
    \caption{Initial conditions for the MHD simulations. The magnetic field is 2 dimensional, with the lines representing the magnetic field lines in the plane. The initial internal energy is a constant.}
    \label{fig:lare_init}
\end{figure}

This system is then allowed to relax with $T_b = 1$ and no dynamic boundary conditions, initially from a state with zero fluid velocity to one with a nonzero vertical `outflow'. Once it has stabilised, horizontal averages of the vertical fluid velocity and density are taken to be used as the initial conditions for the main flux rope simulations. At this point we can introduce the background stratification by reducing $T_b$ (the temperature at the base of the domain relative to the top). Figure \ref{fig:lare_init} shows the initial state of the magnetic field and plasma.

\subsection{Boundary Conditions}

The boundary conditions used for our MHD simulations are almost identical to that of \cite{2023ApJ...955..114R}, in which there is a more detailed discussion on both their numerical implementation and the justification of their ability to emulate physical processes. The magnetic field is constrained such that there is zero perpendicular current on the boundaries, and it is also vertical/radial at the sides of the domain. The internal energy and density have zero-gradient conditions over all of the boundaries except the lower one, whereon the density is held at a constant value one cell within the domain (preventing the system from becoming ever less dense). The boundary conditions on the fluid velocity are more complex, and allow for an equilibirum state with a nonzero vertical flow, roughly representing the effect of the solar wind. 

In addition to these static boundary conditions, we impose dynamic effects at the base of the magnetic field. These boundary flows provide the additional energy to allow for the formation of the magnetic flux ropes. The first of these flows is that of differential rotation, shearing the magnetic arcade in the out-of-plane direction. The velocity profile 
\begin{equation}
V_{\rm {shear}}(x) = -\cos(0.35-x)[0.0031 - 0.041\cos(0.35-x)^2 - 0.031\cos(0.35-x)^4 - V_0] \label{eqn:shearing}
\end{equation} 
is added directly onto the fluid velocity at the lower boundary. This follows from the solar rotation speed profile from \cite{1983ApJ...270..288S}, assuming the centre of our domain is at a latitude of $0.35$ radians. $V_0$ is a reference speed equal to $V_0 = V_{\rm {shear}}(0.35)$. This velocity is added directly onto the plasma velocity field at the base of the domain.

The second dynamic condition we model is an additional magnetic diffusion term acting on the plane of the lower boundary (which is naturally a line in our 2.5D model). This represents the effect of `supergranulation' and serves to bring the magnetic footpoints of the solar arcade together, which (combined with the shearing) allows for reconnection above the polarity inversion line and the formation of a flux rope (\cite{1989ApJ...343..971V}). 

To model this effect we add an extra term $\mathbf{B}_{\rm diff}$ directly to the magnetic field on the lower boundary, where
\begin{align}
    B_{x_\textrm{diff}} &= -\eta_0\frac{\Delta t}{\Delta y} \frac{d}{dx}B_y(x,0) \\
    B_{y_\textrm{diff}} &= \eta_0 \Delta t\frac{d^2}{dx^2}B_y(x,0),
\end{align}
where $\Delta t$ is the timestep and $\Delta y$ is the numerical resolution in the vertical direction. 

\subsection{Model Behaviour and the Effects of Stratification}

\label{sec:delays}

\begin{figure}[ht!]
    \centering
    \includegraphics[width=1.0\textwidth]{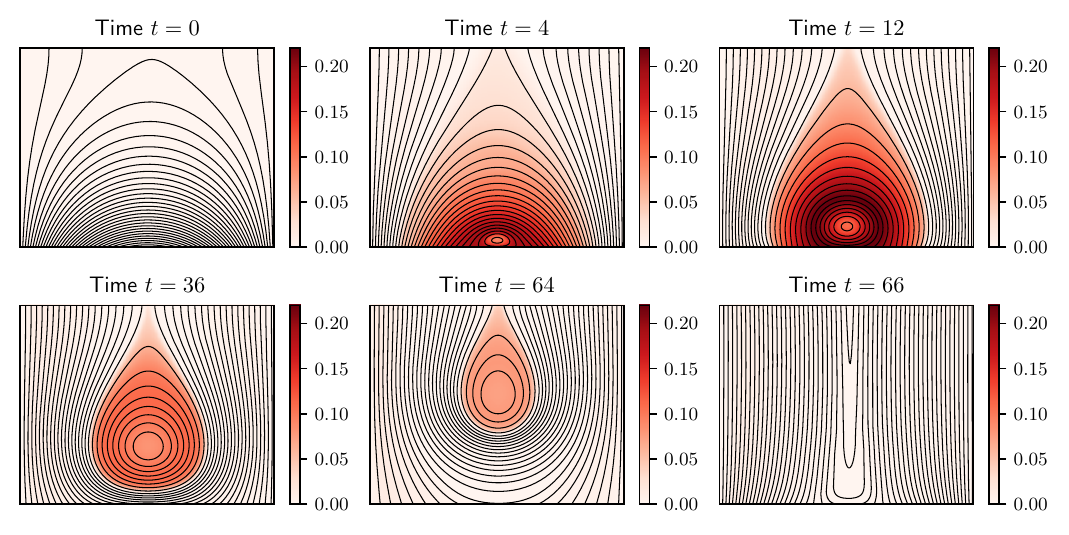}
    \caption{Sequence of snapshots showing a flux rope forming and erupting. The heatmap represents the strength of the out-of-plane magnetic field, and the black lines are the projection of the the magnetic field lines into the plane. The flux rope has already formed by the time of the second snapshot, and erupts shortly after the fifth snapshot at around time $t=65$. In this simulation the stratification factor is $S=150$ and the other variable parameters are $\eta_0 = 2.36 \times 10^{-3}$, $\epsilon_0 = 2$.}
    \label{fig:lare_snaps}
\end{figure}

The first stage in the evolution of our model is the gradual shearing of the magnetic arcade in the out-of-plane ($z$) direction, due to the effect of the differential rotation of the solar surface. The distribution of this out-of-plane field is shown as the heatmap in Figure \ref{fig:lare_snaps}, which shows a sequence of snapshots of the magnetic field evolution.

By time $t=4$, we observe that the flux rope has begun to form -- the rope itself consists of the field lines near the lower centre of the domain that do not meet the boundary. The rope forms as the supergranular diffusion term $\mathbf{B}_{\rm diff}$ essentially brings the magnetic footpoints closer together at the surface until they reconnect, forming a twisted structure. As there is no variance in the $z$ direction, this rope is essentially infinitely long.

As the model evolves, the rope becomes more well-defined, with a larger core and more magnetic flux in the region disconnected from the boundaries. As the flux rope gets larger, it moves upwards, eventually disconnecting itself from the lower boundary (as seen by time $t=36$ in Figure \ref{fig:lare_snaps}). Eventually the system is no longer in quasi-equilibrium and an eruption begins to take place, in this instance at around $t=65$. During this `liftoff phase' the rope moves rapidly upwards through the top boundary of the domain. After this time either a rope reforms or the system remains in a stable, uninteresting state. The time scale and precise nature of the evolution varies considerably based on the parameters $\rho_0, E_0$ and most notably the supergranular diffusion rate $\eta_0$ -- as we can not meaningfully estimate physical values for these parameters we run several sets of simulations over a range of parameters.

\begin{figure}[ht!]
    \centering
    \includegraphics[width=1.0\textwidth]{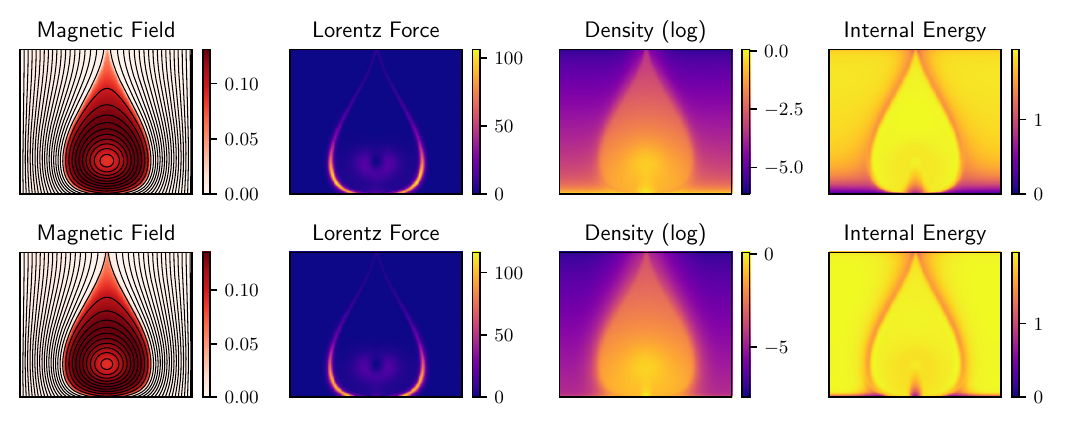}
    \caption{Comparison of an unstratified (top) and stratified (bottom) simulations at the same time during flux rope formation, with all other parameters identical. This is a snapshot at time $t = 16$. The lower pane is the same simulation as shown in Figure \ref{fig:lare_snaps}, and the upper is the unstratified equivalent with $S=1$. The magnetic field is represented as in Figure \ref{fig:lare_snaps}, and the fluid density is plotted on a logarithmic scale as it varies over several orders of magnitude. }
    \label{fig:lare_comp}
\end{figure}

The behaviour of the unstratified model is discussed at length in \cite{2023ApJ...955..114R}, but the focus in this work is on the effects of incorporating a more realistic coronal stratification profile. To this end, we examine $50$ sets of $10$ simulations. Within each set all initial conditions, boundary conditions and almost all parameters are identical -- only the `Stratification Factor' $S = 1/T_b$ is allowed to vary. We choose $1 < S < 150$, geometrically scaled, with the highest value essentially indicating a corona which is $150$ times hotter than the base of the simulation domain (\cite{2021GApFD.115...85M}). We will henceforth refer to the $S=1$ cases as `unstratified'. The other variable parameters are the photospheric diffusion rate, which has values between $1.5 \times 10^{-3} < \eta_0 < 2.5 \times 10^{-3}$ and the initial internal energy density, with $2 < E_0 < 5$.

Figure \ref{fig:lare_comp} compares two simulations within the same set, with $\eta_0 = 2.36 \times 10^{-3}, E_0 = 2$ and at time $t = 24$. At this stage the flux rope has formed but is still far from an eruption. This early in the evolution of the flux rope the magnetic fields look ostensibly similar, irrespective of the stratification. However, we note that the Lorentz Force (which is concentrated primarily around the edge of the flux rope) is marginally stronger in the stratified case. This is in fact generally the case in all our simulations, as will be discussed in Section \ref{sec:lforcemhd}. The characteristics of the plasma in the upper corona are also similar in either case. However, near the lower boundary we observe significant differences in the internal energy and density, as may be expected. Most notably, in the stratified simulations the boundary later is much thinner, with the internal energy and density rising far more quickly moving up through the atmosphere. While this is the case for the open magnetic field, within the flux rope itself the density and temperature are almost identical in both cases. 

\begin{figure}[ht!]
    \centering
    \includegraphics[width=1.0\textwidth]{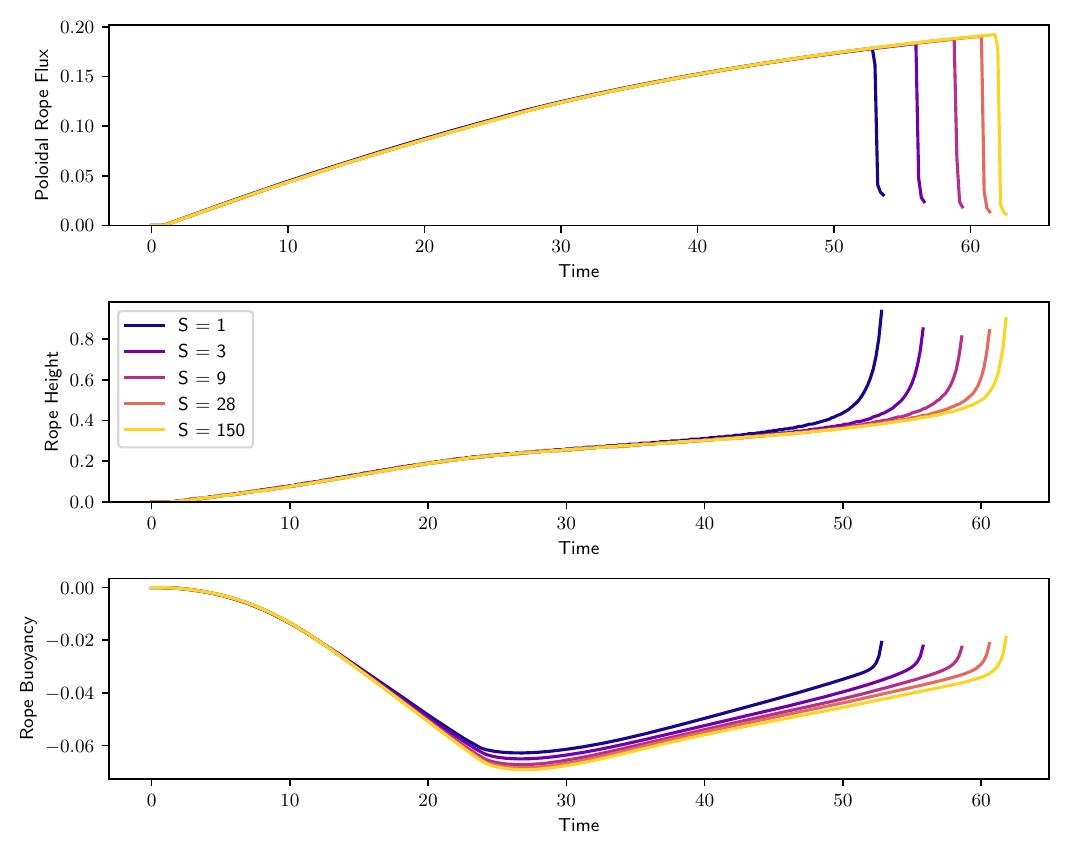}
    \caption{Diagnostic quantities for the set of simulations with $\eta_0 = 2.36 \times 10^{-3}$, $\epsilon_0 = 2$, showing a variety of stratification factors $S$. The top pane plots the poloidal (in-plane) magnetic flux in the rope, the middle pane shows the height of the centre of the rope, and the lower pane plots the `rope buoyancy', defined in Equation \ref{eqn:buoyancy}. }
    \label{fig:diag_set}
\end{figure}

It appears that this density disparity between the background field and the rope itself plays a considerable part in affecting the rope dynamics. In both of the cases shown in Figure \ref{fig:lare_comp}, the rope is more dense than the surrounding open field. This extra `weight', effectively negative buoyancy, serves to hold the rope lower in the atmosphere than the equilibrium position when considering the magnetic field alone. The effect this has can be approximated using a `rope buoyancy' proxy, which we define simply as:
\begin{equation}
    \rm {Rope \, Buoyancy} =  \iint_{\rm Rope \, Area} (\rho_{\rm {back}}(y) -  \rho(x,y))g  \, \,  dx dy, \label{eqn:buoyancy}
\end{equation}
where the `background density' $\rho_{\rm {back}}(y)$ is the vertical density profile in a region near the side of the domain where the field is entirely open, and $g$ is the gravitational acceleration. We plot this buoyancy, along with the height of the centre of the flux rope and the poloidal rope flux (the flux contained within the rope in the in-plane direction) in Figure \ref{fig:diag_set}, for a selection of simulations within the same set.

We note that all the simulations behave very similarly throughout the early stages of the flux rope formation -- and indeed the poloidal rope fluxes are almost identical up to the time of the eruption itself. However, we clearly see that the eruption is delayed when the stratification factor is higher, in this case by up to around $10$ time units. Although the magnetic field is very similar in all cases up to the point at which equilibrium is lost (when the rope height increases more rapidly), the rope buoyancy begins to differ much earlier -- at around $t=20$. This additional negative buoyancy in the stratified cases may influence the point at which the system becomes unstable and the eruption is triggered.

\begin{figure}[ht!]
    \centering
    \includegraphics[width=1.0\textwidth]{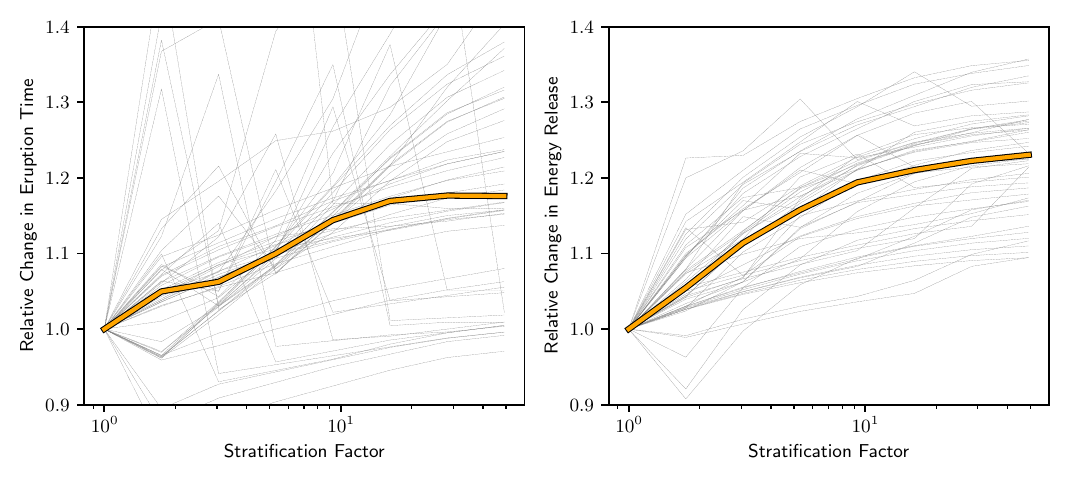}
    \caption{Change in eruption time and energy release with increasing stratification. Each of the thin grey lines represents a sequence of simulations with identical parameters except for the stratification factor. The left pane plots the time taken for an eruption to occur (after the time of the rope's formation) and the right pane plots the magnetic energy released in the eruption. The grey lines are then normalised by the time/energy of the respective unstratified case. The median values are plotted as the thick orange lines.}
    \label{fig:erupt_delays}
\end{figure}

Although the delay in eruptions due to stratification is clear in this particular set of simulations, we need to determine if such a delay is generally the case over a wider range of parameters. Figure \ref{fig:erupt_delays} shows the trends in eruption time and energy release over the entire 50 sets of simulations. Each grey line in the figure represents a set of simulations with all parameters kept constant except the stratification factor, which varies along the $x$ axis. The left panel shows the time taken between the formation of the flux rope and its eruption, determined as the time at which the poloidal rope flux rapidly falls to near zero. Thus, each grey line represents the difference in system behaviour solely due to the introduction of the coronal stratification relative to the unstratified case. These eruption times are normalised relative to the respective unstratified case to better illustrate the trends -- in reality the unstratified eruption times vary significantly.

We observe considerable variation in the effect of the imposed stratification across all 50 sets of runs, which somewhat justifies the use of such a high number. The median value across all sets is shown by the thick orange line. The general trend towards a delay in eruptions is clear, as with the specific case shown in Figure \ref{fig:diag_set}, but this is not the case for all choices of parameters. Indeed, in some cases the introduction of more stratification causes the flux rope to erupt more quickly. This scenario is generally due to there being an earlier failed eruption in the unstratified case, which becomes fully unstable when stratification is added. This effect is not related to the overall trend, and we note that increasing the stratification beyond this point generally causes a delay to the new eruption time as expected.

The median case shows that the highest stratification factor on average delays a flux rope eruption by around $17 \%$, relative to no stratification. Perhaps more notable than the delays in eruptions is the effect on the magnetic energy released in each eruption. It may be expected that the longer the rope takes to form, the more non-potential energy is stored within it and will hence be released, and we observe that this is indeed the case. The relative energy release is plotted on the right-hand pane of Figure \ref{fig:erupt_delays} and shows a more coherent trend than the delay in eruptions alone, with increasing stratification almost always resulting in a higher energy release, with the median increase being around $21 \%$.

\subsection{Effects on the Lorentz Force distribution}
\label{sec:lforcemhd}

Increasing the imposed atmospheric stratification does not appear to influence the magnetic field as severely as the distribution of the plasma (at least while the system is in quasi-equilibrium), but there are nevertheless some notable effects on the magnetic field stricture, especially close to the lower boundary. These can be identified perhaps most clearly in the distribution of the Lorentz Force ($\mathbf{j} \times \mathbf{B}$). When the plasma beta is low, and the magnetic field dominates the dynamics, one would expect the Lorentz Force to be small -- as indeed is a crucial assumption for methods such as magnetofriction to be valid. However, near the lower boundary in our domain the fluid density and pressure are perhaps high enough to require this assumption to be modified. 

Even in 2.5D, The Lorentz Force distribution is not trivial to examine in a meaningful way, as due to the nature of our model there often temporarily exist very small regions with very high forces, which would skew any results. Hence to reduce the effect of these regions we adopt the approach of taking 1D Fourier transforms of the squared $L_2$ norm of the force, taken in the horizontal direction at a given altitude $y$:
\begin{equation}
    \widetilde{L_m}(y,t) = \int \lvert \mathbf{j}(x,y,t) \times \mathbf{B}(x,y,t) \rvert^2 \cos(m \pi x) \, dx, \label{eqn:lorentz1}
\end{equation}
with the cosine term appropriately normalised. We then take an average of this value over all snapshots from all the simulations at a given stratification factor up to the time of eruption: 
\begin{equation}
    L_m(y) = \frac{1}{n_{\rm runs}{n_{\rm snaps}}}\sum_{\rm all \, runs} \sum_t \widetilde{L_m}(y,t). \label{eqn:lorentz2}
\end{equation}

\begin{figure}[ht!]
    \centering
    \includegraphics[width=1.0\textwidth]{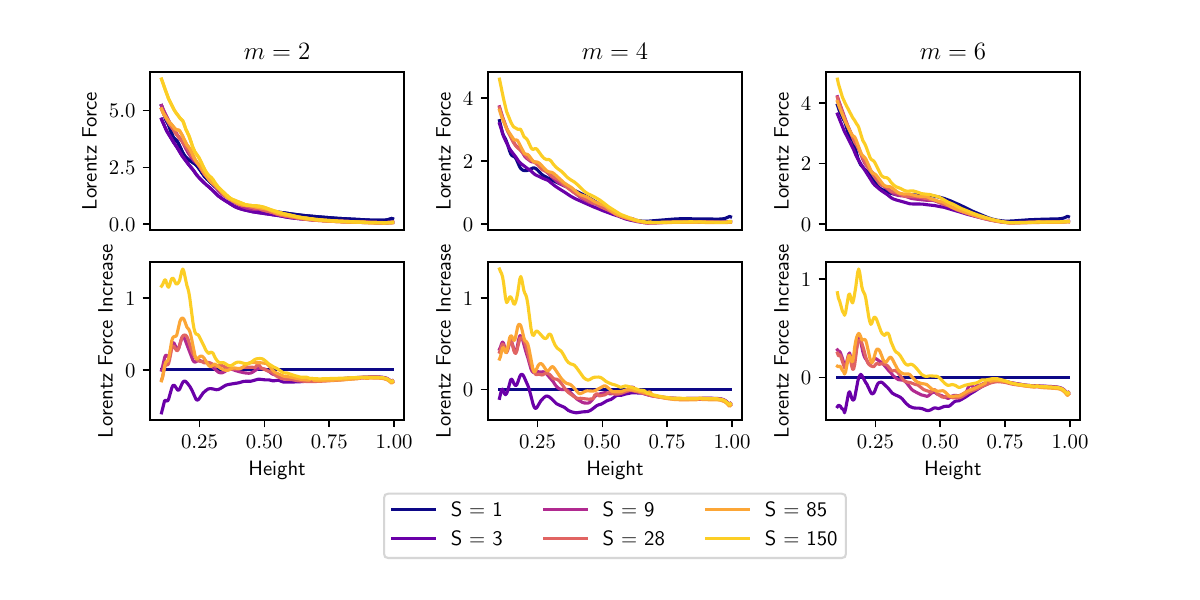}
    \caption{1D Fourier transforms of the absolute Lorentz Force, at a given height $y$. The upper pane shows the absolute values of the quantity $L_m(y)$, for each of the stratification factors averaged over all simulation runs. The lower pane plots these quantities relative to the unstratified case with $S=1$.}
    \label{fig:lorentz_lare}
\end{figure}

Due to the (rough) symmetry of the system, only even Fourier modes have any significance. The first three of these are plotted in Figure \ref{fig:lorentz_lare}.

We see that in all cases the Lorentz Force initially falls roughly exponentially with height, but then above around $y=0.25$ flattens out to decay almost linearly, reducing to zero near the top of the domain. In the case of the higher-order modes this dropoff is more rapid. With increasing stratification the overall forces are generally higher (apart from the first stratification factor plotted here), especially at the base of the domain, where the lowest mode is $25\%$ higher with stratification.

Examining the charge in the force relative to the unstratified cases (the lower panel), we see that stratification only has a significant effect below an altitude of around $0.5$ -- above this height the magnetic field is seemingly unaffected by the nuances of the lower boundary. However, below this altitude we observe quite significant increases in the Lorentz Force, even at altitudes significantly higher than the (very narrow) boundary layer.

This indicates that incorporating even a thin cool boundary later can affect the magnetic field much higher in the domain, increasing the Lorentz Force by up to $25 \%$ relative to the unstratified case. 

\subsection{Discussion}

By modifying the LARE2D code to incorporate the effects of the stratified atmosphere in the solar atmosphere, we have examined the effects that including this improvement has on the formation and eruption of magnetic flux ropes in a simple 2.5D model. We have shown that in general the presence of a significantly more dense and cool layer at the base of the domain (which represents the portion of the solar atmosphere below the transition region) causes a delay in the eruption of flux ropes relative to the `unstratified' case, where there is less of a sharp transition as the system naturally reaches a hydrodynamic equilibrium. This delay in eruptions is accompanied by a proportional increase in magnetic energy per eruption, likely simply as the rope has had longer to form and build up non-potential energy.

We observe that in the early stages of the evolution of our models the magnetic field appears similar irrespective of the degree of stratification -- the main differences are in the properties and distribution of the plasma. However, we observe that generally the Lorentz Force low in the corona is stronger when stratification is imposed, even at altitudes far higher than the transition region. We note that while the density of the flux rope itself is also roughly independent of stratification, the open field surrounding the rope is less dense  in equilibrium when the cooler boundary layer is imposed. This extra negative buoyancy is likely one of the major factors in the observed delay in flux rope eruptions. We hypothesise that this difference may occur as the cool, dense plasma from the boundary layer becomes trapped in the flux rope as it forms, whereas the open field surrounding it is always free to reach a less dense equilibrium state.

Partially as a precursor to attempting to emulate this behaviour using magnetofriction, we analyse the effect that the stratification layer has on the Lorentz Force, by averaging the total force at a given height across all $50$ sets of simulations in our study. We find that above a height of $y = 0.6$ the boundary layer has a negligible effect, but below this height the Lorentz force can increase by up to $25 \%$ relative to the unstratified case, and this increase extends to altitudes far higher than the cool, dense layer itself. The distribution of the Lorentz Force can theoretically be used to judge the success of our modifications to the magnetofrictional model, discussed in the remainder of this paper.

Analysis of the terms in the momentum equation (Equation \ref{eqn:momentum}) can be used to gain additional clarity as to the reasons behind the observed delays in the eruption when the boundary layer is thinner and cooler. Given the system is in quasi-equilibrium as the rope forms, the force from the plasma pressure gradient approximately balances the Lorentz Force and gravity. Imposing stratification does not appear to affect this balance within the rope itself, but does so around its edge -- manifesting as the negative buoyancy described earlier. However, the increase in the Lorentz Force low in the corona as seen in Figures \ref{fig:lare_comp} and \ref{fig:lorentz_lare} is almost completely balanced by more severe plasma pressure gradients low in the corona, and so we propose that this change to the magnetic field structure likely does not affect the overall rope behaviour significantly. An additional effect which may be significant is the nature of the plasma in the open field surrounding the rope -- in the unstratified simulations the plasma pressure (in the positive vertical direction) is higher and extends further up into the corona, which may encourage premature eruptions relative to the more realistic stratified simulations.

\section{The Magnetofrictional Model}

The magnetofrictional (MF) model \cite{1986ApJ...309..383Y}  has long been used in modelling the corona as far less computationally intense alternative to MHD for a variety of applications (eg. \cite{2000ApJ...539..983V, 2012ApJ...757..147C, 2014SoPh..289..631Y}). Unlike in MHD, the fluid equations are disregarded and are replaced by a `magnetofrictional velocity' $\textbf{v}$, which in our `unmodified' model takes the form 
\begin{equation}
    \textbf{v} = \nu_0\frac{\mathbf{j} \times \mathbf{B}}{B^2 + B_0^2\delta e^{-\frac{B^2}{\delta B_0^2}}} + v_{\rm{out}}(y)\textbf{e}_y, \label{eqn:mfv2}
\end{equation}
where $\textbf{j}$ is the current density, $\textbf{B}$ is the magnetic field, $\delta$ is a small parameter chosen to avoid problems at magnetic null points, and $v_{\rm out}$ is an `outflow' function, used to roughly approximate the effect of the solar wind. The constant magnetofrictional relaxation rate $\nu_0$ can be chosen at will and varies in our parameter study. Pure magnetofrictional relaxation (in the absence of an outflow term) serves to return the system to a state with a low Lorentz Force $\mathbf{j} \times \mathbf{B}$, which is a good approximation across much of the corona. However, as discussed in Section \ref{sec:lforcemhd}, this assumption is less valid lower in the atmosphere, where the plasma is more dense. 

A key assumption we make in this work is that this unmodified MF model behaves equivalently to the unstratified MHD model described in the previous section. This was shown in \cite{2023ApJ...955..114R}, which directly discusses the criteria for flux rope eruptions in both MHD and MF and determines several criteria for eruptions which are the same in both cases. It must be noted, however, that in that study the Newton Cooling term (Equation \ref{eqn:coolingterm}) was not included as it is not necessary when the lower boundary is not constrained to be significantly cooler than the rest of the domain.

In Section \ref{sec:mhd} we discussed some of the effects that including this dense layer can have on the eruption of magnetic flux ropes in MHD simulations, finding that on average eruptions are delayed and release more magnetic energy. Moreover, there is an increase in the Lorentz Force low in the domain, even at altitudes well above the transition layer. In this section we propose possible modifications to the existing magnetofrictional model which could emulate these effects to some degree.

The numerical setup is very similar to the MHD equivalent described in the previous chapter, with identical domain sizes, resolutions and magnetic boundary conditions. This is deliberately the case so as to make reasonable comparisons between the two models. The full set of magnetofrictional equations we use is:
\begin{eqnarray}
\frac{\partial \mathbf{B}}{\partial t}&=&-\nabla\times\mathbf{E}\\
\mathbf{E}&=&\eta \mathbf{j} - \mathbf{v}\times\mathbf{B} \\
\mathbf{j}&=& \nabla\times\mathbf{B}\\
\mathbf{v} &=& \nu_0\frac{\mathbf{j} \times \mathbf{B}}{B^2 + B_0^2\delta e^{-\frac{B^2}{\delta B_0^2}}} + v_{\rm{out}}(y)\textbf{e}_y,
\end{eqnarray}
where the constants not introduced in Equation \ref{eqn:mfv2} have the same meaning as in the MHD model. 

The photospheric shearing rate is applied directly to the magnetofrictional velocity as before. The outflow term $v_{\rm{out}}(y)\textbf{e}_y$ takes the profile 
$v_{\rm {out}}(y) = y^4$, and is added on to the velocity field only in the region where the magnetic field is open -- determined as the area consisting of field lines which intersect the top boundary. This new approach is used to encourage a similar current density distribution to that seen in the MHD models, although it is only successful to a small degree. The supergranular diffusion term replaces the electric field on the lower boundary at each timestep:
\begin{equation}
{\bf E}(x,0) = \eta_0\frac{\partial B_y(x,0)}{\partial x}{\bf e}_z,
\end{equation} 
essentially increasing the magnetic diffusion rate in the plane/line of the solar surface.

As in the MHD simulations, we undertake a parameter study to observe the effect of our modifications on a variety of flux ropes, varying the supergranular diffusion rate $\eta_0$ as before but now also varying the magnetofrictional relaxation rate $\nu_0$, which is difficult to assign a true physical value. The ranges we choose for these parameters are $7.5 \times 10^{-4} < \eta_0 < 1.25 \times 10^{-3}$ and $0.025 < \nu_0 < 0.1$.

We now consider modifications to this model which we propose could better emulate the effects of atmospheric stratification.

\subsection{Rope Weight}

While discussing the influence of the lower boundary layer in Section \ref{sec:mhd} we note that when the imposed stratification is more prevalent (ie. the lower boundary is cooler relative to the corona), the flux rope which forms has a more dense core relative to the background plasma. We hypothesise that this is due to the more dense plasma at the base of the domain remaining trapped in the rope as it forms, whereas in the surrounding open field the plasma is free to flow outwards and equilibriate at a lower density. Although during the formation of the rope the dynamics are very similar, this extra weight eventually serves to delay the liftoff phase and eruption of the flux rope.

In our simple MF model it is not particularly difficult to emulate this effect, by adding a negative vertical velocity at each timestep to the flux rope itself (the region which remains more dense in the MHD simulations). This region is nicely defined in our 2.5D model as it comprises the field lines which never touch the outer boundary. In more complex 3D models an equivalent modification would be far more difficult to achieve, although it could theoretically be possible by determining a threshold in a topological quantity such as the field line helicity \cite{1988A&A...201..355B} or twist \cite{2006JPhA...39.8321B}, above which the weight mask is applied. 

Defining the downward velocity due to the `weight' as $v_{\rm w}\textbf{e}_y$, the magnetofrictional velocity is then, explicitly

\begin{equation}
    \textbf{v} = \nu_0\frac{\mathbf{j} \times \mathbf{B}}{B^2 + B_0^2\delta e^{-\frac{B^2}{\delta B_0^2}}} + v_{\rm{out}}(y)\textbf{e}_y - v_{\rm w}\textbf{e}_y, \label{eqn:mfv}
\end{equation}
where the outflow term is applied only to the region comprising open field lines, and the weight term is applied only to the flux rope itself. We find, as expected, that eruptions are delayed by the addition of $v_{\rm w}$, and values of $v_{\rm w}  \gtrapprox 0.05$ prevent eruptions entirely.

\subsection{Pressure Current}

\begin{figure}[ht!]
    \centering  \includegraphics[width=1.0\textwidth]{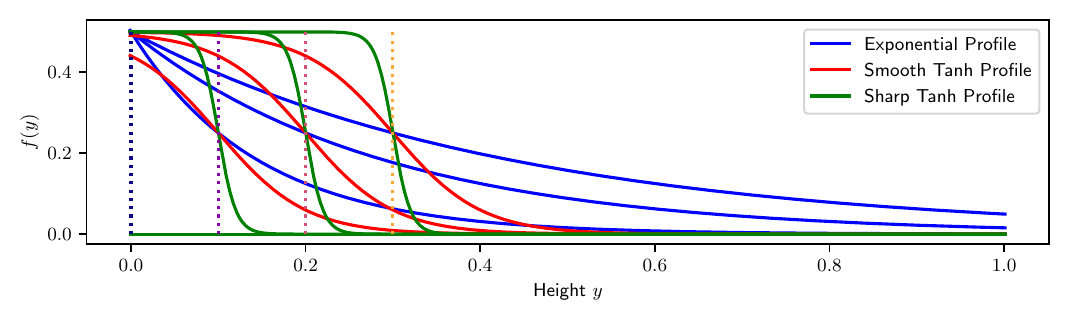}
    \caption{Plot of the three forms of the pressure functions we consider, with $f_{\rm exp}(y)$ in blue, $f_{\rm smooth}(y)$ in red and $f_{\rm sharp}(y)$ in green. The three nonzero profiles of each form correspond to the cutoff height $y_0$ at which the pressure function decays to half of its value at the base of the domain.}
    \label{fig:pressure_functions}
\end{figure}

Although possible in our model, the addition of the `rope weight' as described here would generally be more difficult (especially in 3D) due to the somewhat loose definition of what comprises the flux rope itself. It would thus be preferable to modify the model in a way that does not rely on such a definition. One suitable alternative we propose is the addition of a `pressure current' $\mathbf{j}_p$, which depends only on the magnetic field and a specified function $f(y)$.

The motivation here is that low in the corona we do not wish the system to return to a force-free state as quickly as it does higher up. Whereas the magnetofrictional model will (in the absence of any additional terms in the velocity equation) tend to an equilibirum with $\textbf{v} = 0$ and hence $\textbf{j} \times \textbf{B}  = 0$, we wish the system to instead be in equilibrium when the Lorentz Force is balanced by a specified plasma pressure gradient and gravity. In such an equilibrium \cite{2019SoPh..294..171N} propose that when
\begin{equation}
    \mathbf{j} \times \mathbf{B} - \nabla P - \rho g \mathbf{e}_y= 0,
\end{equation}
with $P$ the plasma pressure, $\rho$ the plasma density and $g$ gravitational acceleration, the current density would take the form
\begin{equation}
    \mathbf{j} = \alpha \mathbf{B} + \nabla \times (f(y) B_y \mathbf{e}_y), \label{eqn:forcenotfree}
\end{equation}
where $\alpha$ is a force-free parameter, and $f(y)$ is the previously mentioned `pressure function'. Clearly if $f(y) = 0$ we recover a force-free field as in a standard magnetofrictional equilibrium, as is the case sufficiently high in the corona.

We wish to modify the magnetofrictional method such that equilibrium solutions take the form of Equation \ref{eqn:forcenotfree}. We can achieve this simply by defining the `pressure current' as $\mathbf{j}_p = \nabla \times (f(y) B_y \mathbf{e}_y)$, and subtracting this from the usual current density, so that the magnetofrictional velocity is now 
\begin{equation}
    \widetilde{\mathbf{v}} = \frac{\nu_0}{B^2 + B_0^2\delta e^{-\frac{B^2}{\delta B_0^2}}}(\mathbf{j} \times \mathbf{B} - \mathbf{j}_p \times  \mathbf{B}).
\end{equation}
It must be noted that in the absence of dynamic boundaries this modified system will not necessarily monotonically converge to the equilibirum from Equation \ref{eqn:forcenotfree}, although experimentally it does appear that the system tends towards a similar state, as we will discuss below when examining the Lorentz Force distribution of our simulations.

We consider three forms for the pressure function $f(y)$. The first is the `exponential' profile
\begin{equation}
    f_{\rm exp}(y) = \frac{1}{2} e^{-y/b},
\end{equation}
with $b = y_0/\ln(2)$, such that $f(0) = 0.5$ and $f(y_0) = 0.25$ for a given cutoff height $y_0$. The use of a function of this form in magnetohydrostatic (MHS) models of the corona dates back to \cite{1991ApJ...370..427L}, who find a semi-analytic solution where the vertical eigenmodes are Bessel functions. This function has also been used extensively for MHD modelling purposes (eg. \cite{1998SoPh..183..369A}, \cite{2015ApJ...815...10W}). We see in Figure \ref{fig:lorentz_lare} that the increase in Lorentz Force due to MHD stratification decays roughly exponentially at low altitudes, which is in itself good motivation to consider a function of this form.

The second and third forms we consider are `tanh' profiles:
\begin{equation}
    f_{\rm smooth}(y) = 0.25 \left[ 1 - \tanh \left( \frac{y-y_0}{0.1} \right) \right],
\end{equation}
and
\begin{equation}
    f_{\rm sharp}(y) = 0.25 \left[ 1 - \tanh \left( \frac{y-y_0}{0.02} \right) \right].
\end{equation}
The only difference between these being the denominator in the $\tanh$ function, which controls how quickly the pressure current falls to zero. These three forms of $f(y)$ are plotted in Figure \ref{fig:pressure_functions}, noting that the cutoff height $y_0$ is the point at which the pressure function falls to half its maximum value in every case.

The $\tanh$ pressure function was introduced in an MHS context by \cite{2019SoPh..294..171N} and have several advantages over the exponential equivalent, namely that there are more degrees of freedom and that these functions result in MHS fields which appear to match observations more closely than an exponential decay. In particular, it is noted that the magnetic field lines become more vertical low in the corona than the $f(y) = 0$ equivalent, matching certain observations of active regions. 

We have chosen $f(0) \approx 0.5$ for consistency between the three forms of $f(y)$ -- at values much higher than this the magnetofrictional scheme is more likely to become numerically unstable. Naturally it would be possible to scale any of these functions by an overall factor, but our interest here lies mainly in the effects of the shape of the functions and their decay rate, rather than their overall magnitude.
 
\subsection{The Effects of the Magnetofrictional Modifications}

Similarly to the MHD parameter study described in Section \ref{sec:mhd}, we simulate the formation and eruption of magnetic flux ropes over a range of parameters ($\eta_0, \nu_0$), establishing the `unstratified' base behaviour with $f(y) \equiv 0$ for each of $50$ parameter sets. Within each set we then introduce either of the modifications: adding the `rope weight' or a `pressure function' as appropriate to examine the effect this has on the flux rope behaviour. In this study we vary the rope weight parameter between $0 < v_w < 0.03$. While using the pressure functions we vary the cutoff height in the range $0 < y_0 < 0.3$, as shown in Figure \ref{fig:pressure_functions}.

\begin{figure}[ht!]
    \centering  \includegraphics[width=1.0\textwidth]{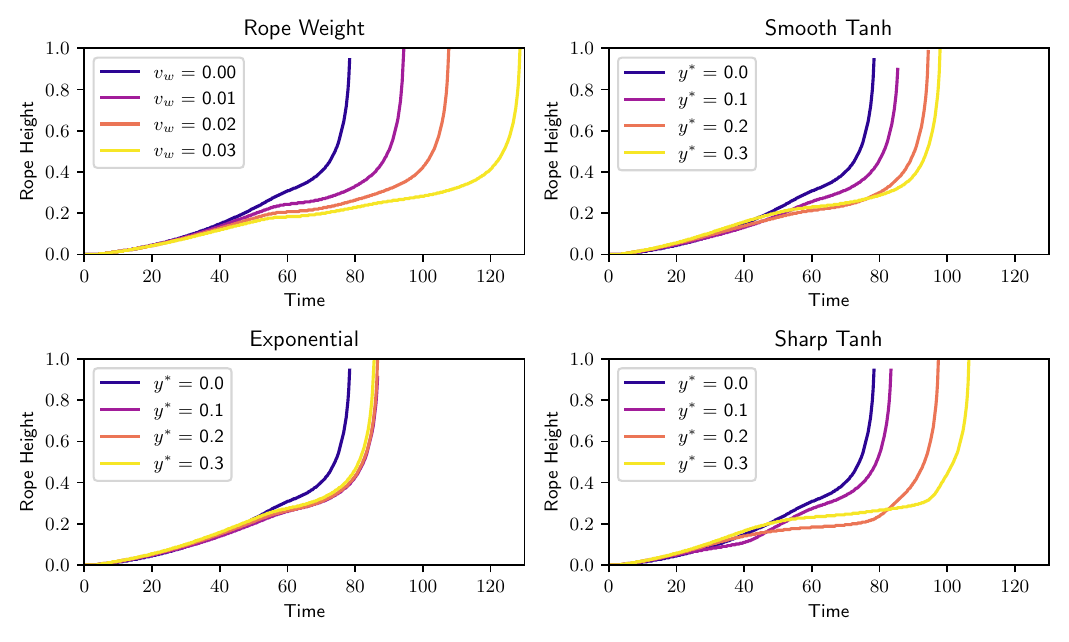}
    \caption{Plots of the height of the centre of the flux rope from the simulations with $\eta_0 = 0.001, \nu_0 = 0.06$, showing the delay in eruption due to the effect of the three modifications to magnetofriction that we test. The runs are colour coded based on the modifcation parameters used in each case.}
    \label{fig:diag_mf}
\end{figure}

In all of the four modification scenarios we observe, on average, a delay in the flux rope eruptions. For a single set of parameters, Figure \ref{fig:diag_mf} plots the heights of the flux rope centres up to the time of the first eruption. Adding the rope weight modification effectively delays eruptions in a manner ostensibly very similar to the `ground truth' LARE simulations shown in Figure \ref{fig:diag_set}, with the rope initially rising at roughly the same rate independently of $v_w$ and then the lighter ropes rising more quickly. Increasing $v_w$ beyond this range would likely prevent an eruption entirely. Adding the exponential pressure function produced perhaps the most unexpected results: for a low cutoff height $y^*$ the eruptions were delayed, but as the cutoff height increased further beyond around $y^* = 0.1$, this effect lessened and did not further inhibit the rise of the rope.

The addition of the tanh pressure functions had more of an effect on the dynamics, with notable differences manifesting themselves earlier on in the simulations than the MHD `ground truth'. This is particularly the case for the `sharp tanh' profiles, where the introduction of the pressure current caused significant changes to the behaviour of the rope from as early as $t=20$. Unlike when adding `rope weight' the addition of the pressure function does not necessary result in a lower equilbirium position at all times. Indeed we note that the ropes with higher `sharp' profiles initially rise more quickly than the unmodified equivalents, although they do ultimately take longer to erupt. 

\begin{figure}[ht!]
    \centering
    \includegraphics[width=1.0\textwidth]{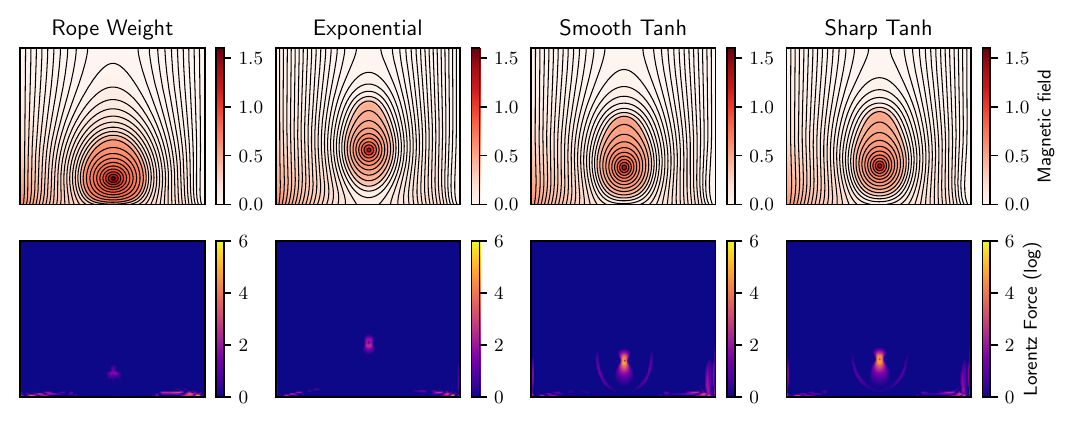}
    \caption{Comparison of the four modifications to the magnetofrictional model, showing the magnetic field (top panes) and the logarithm of the Lorentz Force (bottom panes). The snapshots are taken at time $t=75$ from simulations with $\eta_0 = 0.001$ and $\nu_0 = 0.06$. The cutoff height for the respective pressure functions is $y^* = 0.3$ and the rope weight parameter is $v_w = 0.03$.}
    \label{fig:mf_comp}
\end{figure}

As an indication of how these modifcations affect the flux rope structure itself, Figure \ref{fig:mf_comp} presents snapshots of the rope with each of the modifications in turn, each at the same time $t=75$ and with all other parameters kept the same. The changes to the magnetic field structure are subtle but not insignificant. When adding the weight term the flux rope is in equilibirum lower in the atmosphere, and is also more `bottom heavy', with more area below the centre the rope than above. This shape appears to be the closest to the MHD `ground truth' of the options we study. The simulation with an exponential function is the closest in time to erupting, and it sits higher and is more stretched vertically. The two tanh functions are almost indistinguishable from each other and qualitatively lie somewhere in between the exponential and rope weight cases. 

Also of note is the distribution of the Lorentz Force, plotted in the lower pane of Figure \ref{fig:mf_comp}. Apart from some areas of high force near the lower boundary (which are merely an artifact of how the boundary conditions are applied), most of the Lorentz Force is concentrated within the core of the rope itself, with some also found around the edge of the rope at the boundary between the open and closed field (this is more clear in the MHD simulations shown in Figure \ref{fig:lare_comp}). Notably, the addition of the tanh pressure function increases the Lorentz Force in the rope core considerably -- far more than the exponential function and the rope weight equivalent. This increase is likely the ultimate cause of the delay in eruptions, given that unlike in the MHD case there are no plasma pressure effects to consider in the momentum equation.

\begin{figure}[ht!]
    \centering
    \includegraphics[width=1.0\textwidth]{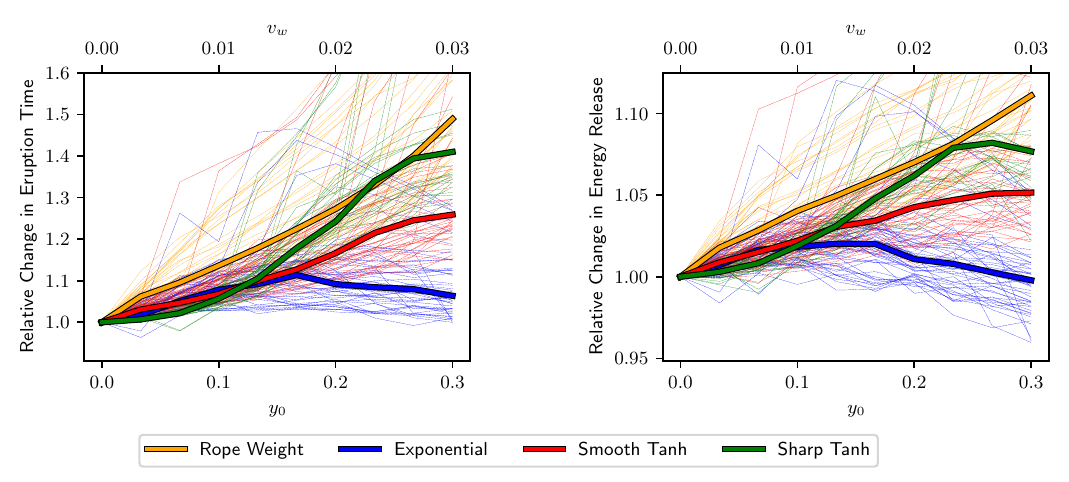}
    \caption{Plots of the delays in the time of flux rope eruptions and the respective releases in magnetic energy, as in Figure \ref{fig:erupt_delays}. All three modifications to the MF model are plotted, and are coloured respectively. All individual simulation sets are plotted as the thin lines, with the median values plotted as the thick outlined lines.}
    \label{fig:delays_mf}
\end{figure}

Similarly to the MHD simulations in Section \ref{sec:mhd} we test if these conclusions are generally true across the entire parameter range of $50$ simulation sets. To this end, the relative eruption times and energy releases are shown in Figure \ref{fig:delays_mf}. The premise of this plot is the same as Figure \ref{fig:erupt_delays}, but the data are now coloured based on the respective modification to the MF model. 

Adding the rope weight modification always delays eruptions, with our maximal parameter value $v_w$ sometimes preventing eruptions entirely within the timeframe of our simulations. To achieve the $\approx 20\%$ delay in eruptions seen in the MHD simulations a value of $v_w \approx 0.015$ is required -- although at this value the average increase in eruption energy release is more modest at $5\%$, which is much less than the $\approx 20\%$ seen in the stratified MHD simulations. 

The addition of the exponential pressure function has the least effect on eruption timing or energy release. Indeed in fact as the cutoff height $y^*$ increases beyond around $y^* \approx 0.15$ the eruptions are delayed less than for lower cutoff heights, so indeed it is seemingly the profile and decay rate of the pressure function that ultimately determines the effect on the system rather than merely its magnitude. The average energy release follows a similar pattern.

The tanh pressure functions, in contrast, delay eruptions very effectively when $y^*$ is sufficiently large, with the median flux rope lasting $40\%$ longer when the sharp tanh pressure function is imposed with $y^* = 0.3$. However, even in this extreme case the magnetic energy released still does not increase by as much as the stratified MHD simulations, with still less than a $10\%$ increase relative to the unstratified and unmodified simulations. The general trend of both sharp and gentle tanh functions are similar, but the sharp profile is far more effective, indicating that a steep decay in $f(y)$ is crucial for this approach to be effective.

It is notable that even in this simple model, with a relatively small range of parameters, that each simulation set behaves very differently, as evidenced by the large variation in the thin coloured lines in Figure \ref{fig:delays_mf} -- only by taking the median value over a large range can we gain any insight. Thus it must be emphasised that in general drawing significant conclusions from simulations which assume a single set of parameters may often be misguided.

\begin{figure}[ht!]
    \centering
    \includegraphics[width=1.0\textwidth]{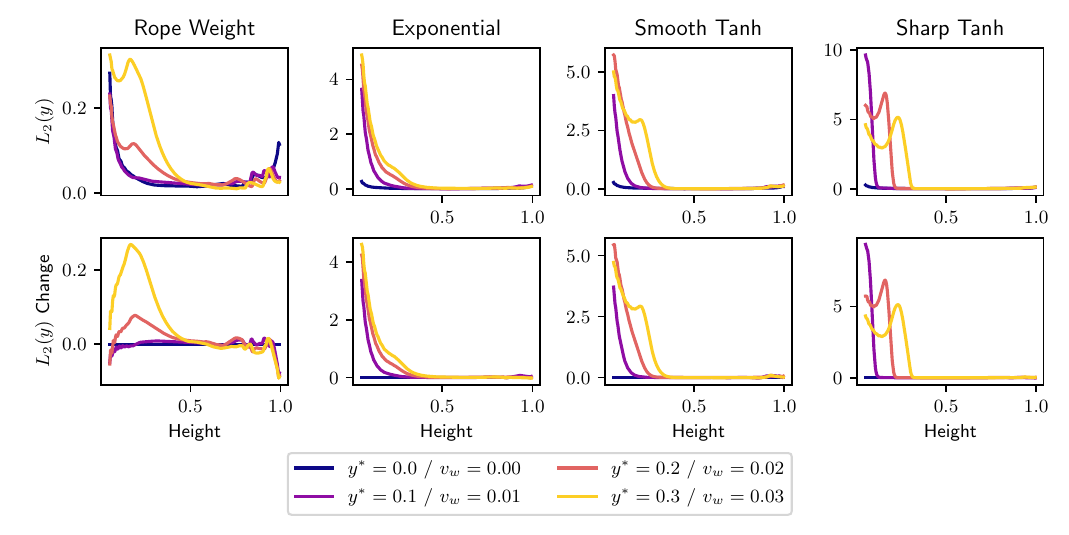}
    \caption{Plots of the Fourier Transform $L_2(y)$ of the average Lorentz Force across all 50 magnetofrictional simulation sets, for each of the three modifications in turn. This plot only includes the first significant Fourier mode for each case: $m=2$ from Equation \ref{eqn:lorentz2}. The lower panels plot the difference between the modified and unmodified cases, similarly to Figure \ref{fig:lorentz_lare}.}
    \label{fig:lorentz_mf}
\end{figure}

\subsection{Magnetofrictional Lorentz Force Distribution}

It remains to compare the Lorentz Force distribution with the inclusion of these modifications, and compare against the MHD equivalents. We once again calculate the 1D Fourier Transform of the Lorentz Force with height, using Equations \ref{eqn:lorentz1} and \ref{eqn:lorentz2}. The results for the first significant Fourier mode ($m=2$) are plotted in Figure \ref{fig:lorentz_mf}. As discussed in Section \ref{sec:lforcemhd}, it is likely that although adding background stratification does increase the Lorentz Force at low altitudes, it is unclear whether this directly affects the overall system behaviour. In general the unmodified MF cases very much underestimate the Lorentz Force at low altitudes, as can be seen by comparing Figure \ref{fig:lorentz_lare} and Figure \ref{fig:lorentz_mf}.

We observe that although adding the weight of the rope could theoretically delay flux rope eruptions indefinitely, and appeared to show more MHD-like behaviour than the pressure functions, the overall magnitude of the Lorentz Force remains roughly unchanged at the lower boundary. The simulations with the highest rope weight parameter did exhibit a slower decay in the Force than the unmodified case, which is likely due to the presence of a rope existing at higher altitudes for longer in this case. Adding the exponential pressure function predictably resulted in an exponential profile to the Lorentz Force decay -- very similarly to that observed in the MHD simulations and of the correct order of magnitude (around $5$ units at the lower boundary).

The smooth tanh pressure function is not too dissimilar, although the decay of the Lorentz Force with height is a little faster. Moreover, the absolute magnitude of the force near the lower boundary is very similar to the MHD simulations, at around $5$ units. Unlike the exponential pressure function, we also observe a delay in eruption times commensurate with adding stratification to the MHD model, indicating that this function may be the best of both worlds. The sharp tanh pressure function had the most extreme effect on the Lorentz Force, which (as expected) falls rapidly at the cutoff height $y^*$. Both the cutoff height and the sharpness of the cutoff also significantly effect the maximum force on the lower boundary, despite the maximum value of the pressure function ($f(0)$) being identical in every case. 

\subsection{Discussion}

We have performed a study using the magnetofrictional (MF) model to determine how a series of new additions to the model can be used to better emulate full MHD simulations of magnetic flux rope eruptions. The principle of the MF model is that the system exists in a series of quasi-equilibria with a low Lorentz Force, which is a good approximation across much of the corona, but not necessarily near the solar surface. 

We introduce modifications to the model to address both this shortcoming, and to better represent the effects of the atmospheric stratification discussed previously in the paper. The addition of a `pressure current' to the region low in the corona improves upon the model by increasing the Lorentz Force at these low altitudes. We evaluate three forms of this current distribution, and determine that the decay rate of the corresponding `pressure function' makes a large difference to the model behaviour. A sharp cutoff height at which the pressure function falls quickly to zero delays the formation and eruption of flux ropes. Of the three pressure profiles considered, an exponential decay matches the MHD Lorentz Force distribution well, and a `tanh' decay delays flux rope eruptions in a manner similar to the introduction of stratification to the MHD model. A `smooth tanh' profile as seen in Figure \ref{fig:pressure_functions} seems to be a good compromise, with both a realistic Lorentz Force distribution and flux rope behaviour.

An alternative modification we consider is the addition of a `rope weight', whereby we simply add a negative velocity to the region comprising the flux rope. This modification delays eruptions in a manner physically very similar to the effect of stratification in MHD, but does not improve upon the unrealistic lack of current low in the atmosphere. This modification could of course be combined with a pressure current to address this problem, emulating the effect of MHD stratification with both an increase in low-altitude Lorentz Force and a suitable delay in eruptions due to the negative buoyancy of the flux rope itself.

\section{Conclusions}

The aims of this paper have been twofold. In the first section we described how the addition of a cool, dense boundary layer affects the formation and eruption of magnetic flux ropes in an MHD model. This boundary layer represents the presence of the chromosphere, which is frequently ignored or underrepresented in coronal models. This has the potential to be problematic given that the magnetic field structure is often informed by observations from the photosphere -- below this layer. We find that even though the chromosphere only makes up less than $1\%$ of our computational domain, the effects are significant: flux rope eruptions are generally delayed by up to $20\%$ (relative to the time of their formation), with a corresponding increase in the amount of magnetic energy released in eruptions.

We also find that adding the realistic background stratification increases the average Lorentz Force at altitudes far higher than the boundary layer. However, this increased force appears to be mainly balanced by larger pressure gradients and so is likely not the dominant factor in the altered flux rope behaviour. We instead hypothesise that the dense material from the boundary layer becomes trapped within the flux rope, reducing its buoyancy relative to the open field background. This principle would theoretically apply to any similar eruptive scenario where there exists a region of twisted magnetic field capable of trapping dense plasma for some time.

Although ideally MHD would always be used for such models, due to its more intense computational nature it is still unsuitable for some scenarios (eg. real-time global modelling of the corona), and so we wish to see if these findings can be used to improve upon a frequently used and much simpler model -- magnetofriction. In the second section of the paper we discuss potential modifications to the magnetofrictional model which can more accurately represent the true nature of the lower corona, where the `force-free' assumption is less valid. Motivated by existing work on MHS fields, our addition of a `pressure current' achieves this to a degree, resulting in Lorentz Forces in the lower corona which are very similar to the MHD equivalents. We show that the decay rate with altitude of these pressure currents makes a large difference to the flux rope dynamics -- with a steep dropoff at a given altitude delaying eruptions very effectively. The most successful of the decay profiles we considered was the `smooth tanh' option, which combined a realistic Lorentz Force distribution with a reasonable delay in eruptions. The more commonly-used exponential decay did not perform well in comparison, with barely any effect on the overall flux rope dynamics and no average increase in the energy release during eruptions.

We also consider the addition of a `rope weight' modification, introducing an element of the fluid dynamics which is otherwise absent from the magnetofrictional model. This is informed by our observation that the introduction of a dense boundary layer in MHD increases the density of a flux rope relative to the background field. By applying an additional negative velocity to the region within the rope (which is easy to define in our 2.5D models) we can emulate this effect very effectively. The rope weight addition can theoretically be combined with a suitable pressure function to produce a more realistic force distribution if required. We note that although our ropes are well-defined as the region with closed magnetic field lines, in a 3D system determining the region to which the weight is applied would require the use of a proxy such as field line helicity or twist, which would increase the computational complexity somewhat -- but still far less so than using full MHD.

Although the simple 2.5D models we have focused on in this paper are only suitable for very limited applications, the modifications that we have discussed have the potential to be applied to a variety of more complex coronal models. By showing that the precise nature of the lower boundary layer can affect dynamics at far higher altitudes we have identified the importance of modelling this as accurately as possible in MHD simulations. In future we intend to test whether the delays in eruptions we observe also occur in full 3D simulations, both in a coronal jet model (eg. \cite{2009ApJ...691...61P}) and more complex 3D flux rope models. We expect that when the magnetic field is twisted due to boundary motions the same effects are likely to occur as in our simulations here, but it is not clear whether this would also be the case for models in which pre-twisted flux emerges from the photosphere (eg. \cite{2021NatCo..12.6621M}). 

More broadly, the effects of enforcing atmospheric stratification in MHD models is reasonably well-studied, and so it perhaps our modifications to the magnetofrictional model which may be of more significance.

The addition of the `pressure current' to existing 3D magnetofrictional models both globally and for a single active region (eg. \cite{2024ApJ...961L...3A}, \cite{2014ApJ...782...71G}, \cite{2012ApJ...757..147C}) would be a relatively simple process, although determining the magnitude and precise profile of this term may pose challenging. For simulations of single events or active regions, it may be possible to apply a similar approach to the MHS simulations of \cite{2023SoPh..298....3W}, whereby the optimum parameters are determined algorithmically by comparison with emissions from coronal loops. However, this is likely impossible for global simulations and the standard trial-and-error approaches to determining ideal magnetofrictional parameters (such as the relaxation and hyperdiffusion rates) would have to be used. Once achieved, it would  would be desirable to evaluate the effect this has on eruptive events -- specifically to examine the Lorentz Force increase low in the corona and to examine any changes to the frequency and energy release of these events. Over long time scales, such as those used in large-scale flux-transport models (eg.  \cite{2014SoPh..289..631Y}), the effects on the timing of eruptions and their respective energy release could build to be quite significant, and potentially have implications for the space-weather forecasting that such simulations can be used to inform.  The addition of the `rope weight' modification would also be possible but more difficult to implement due to the more vague definition of what constitutes the rope when working in three dimensions.

We hope that this study can be used to inform the community of some of the effects that commonly-used assumptions in coronal modelling can have. Of particular note we find that incorporating an accurate boundary layer to represent the chromosphere can have a significant effect on simulations using magnetic field data from photospheric observations, even if this layer comprises a very small proportion of the computational domain.



\bibliography{main}{}

\begin{thebibliography}{}
\expandafter\ifx\csname natexlab\endcsname\relax\def\natexlab#1{#1}\fi
\providecommand{\url}[1]{\href{#1}{#1}}
\providecommand{\dodoi}[1]{doi:~\href{http://doi.org/#1}{\nolinkurl{#1}}}
\providecommand{\doeprint}[1]{\href{http://ascl.net/#1}{\nolinkurl{http://ascl.net/#1}}}
\providecommand{\doarXiv}[1]{\href{https://arxiv.org/abs/#1}{\nolinkurl{https://arxiv.org/abs/#1}}}

\bibitem[{{Arber} {et~al.}(2001){Arber}, {Longbottom}, {Gerrard}, \& {Milne}}]{2001JCoPh.171..151A}
{Arber}, T.~D., {Longbottom}, A.~W., {Gerrard}, C.~L., \& {Milne}, A.~M. 2001, Journal of Computational Physics, 171, 151, \dodoi{10.1006/jcph.2001.6780}

\bibitem[{{Aslanyan} {et~al.}(2024){Aslanyan}, {Meyer}, {Scott}, \& {Yeates}}]{2024ApJ...961L...3A}
{Aslanyan}, V., {Meyer}, K.~A., {Scott}, R.~B., \& {Yeates}, A.~R. 2024, \apjl, 961, L3, \dodoi{10.3847/2041-8213/ad1934}

\bibitem[{{Aulanier} {et~al.}(1998){Aulanier}, {D{\'e}moulin}, {Schmieder}, {Fang}, \& {Tang}}]{1998SoPh..183..369A}
{Aulanier}, G., {D{\'e}moulin}, P., {Schmieder}, B., {Fang}, C., \& {Tang}, Y.~H. 1998, \solphys, 183, 369, \dodoi{10.1023/A:1005003426798}

\bibitem[{{Berger}(1988)}]{1988A&A...201..355B}
{Berger}, M.~A. 1988, \aap, 201, 355

\bibitem[{{Berger} \& {Prior}(2006)}]{2006JPhA...39.8321B}
{Berger}, M.~A., \& {Prior}, C. 2006, Journal of Physics A Mathematical General, 39, 8321, \dodoi{10.1088/0305-4470/39/26/005}

\bibitem[{{Chen} {et~al.}(2014){Chen}, {Peter}, {Bingert}, \& {Cheung}}]{chen2014model}
{Chen}, F., {Peter}, H., {Bingert}, S., \& {Cheung}, M.~C.~M. 2014, \aap, 564, A12, \dodoi{10.1051/0004-6361/201322859}

\bibitem[{{Chen} {et~al.}(2022){Chen}, {Rempel}, \& {Fan}}]{chen2022comprehensive}
{Chen}, F., {Rempel}, M., \& {Fan}, Y. 2022, \apj, 937, 91, \dodoi{10.3847/1538-4357/ac8f95}

\bibitem[{{Cheung} \& {DeRosa}(2012)}]{2012ApJ...757..147C}
{Cheung}, M. C.~M., \& {DeRosa}, M.~L. 2012, ApJ, 757, 147, \dodoi{10.1088/0004-637X/757/2/147}

\bibitem[{{Cheung} \& {Isobe}(2014)}]{cheung2014flux}
{Cheung}, M. C.~M., \& {Isobe}, H. 2014, Living Reviews in Solar Physics, 11, 3, \dodoi{10.12942/lrsp-2014-3}

\bibitem[{{Doyle} {et~al.}(2019){Doyle}, {Wyper}, {Scullion}, {McLaughlin}, {Ramsay}, \& {Doyle}}]{doyle2019observations}
{Doyle}, L., {Wyper}, P.~F., {Scullion}, E., {et~al.} 2019, \apj, 887, 246, \dodoi{10.3847/1538-4357/ab5d39}

\bibitem[{{Gibb} {et~al.}(2014){Gibb}, {Mackay}, {Green}, \& {Meyer}}]{2014ApJ...782...71G}
{Gibb}, G.~P.~S., {Mackay}, D.~H., {Green}, L.~M., \& {Meyer}, K.~A. 2014, \apj, 782, 71, \dodoi{10.1088/0004-637X/782/2/71}

\bibitem[{{Guo} {et~al.}(2016){Guo}, {Xia}, \& {Keppens}}]{guo2016magneto}
{Guo}, Y., {Xia}, C., \& {Keppens}, R. 2016, \apj, 828, 83, \dodoi{10.3847/0004-637X/828/2/83}

\bibitem[{{Guo} {et~al.}(2019){Guo}, {Xia}, {Keppens}, {Ding}, \& {Chen}}]{guo2019solar}
{Guo}, Y., {Xia}, C., {Keppens}, R., {Ding}, M.~D., \& {Chen}, P.~F. 2019, \apjl, 870, L21, \dodoi{10.3847/2041-8213/aafabf}

\bibitem[{{Inoue} {et~al.}(2023){Inoue}, {Hayashi}, {Miyoshi}, {Jing}, \& {Wang}}]{inoue2023comparative}
{Inoue}, S., {Hayashi}, K., {Miyoshi}, T., {Jing}, J., \& {Wang}, H. 2023, \apjl, 944, L44, \dodoi{10.3847/2041-8213/acb7f4}

\bibitem[{Inoue {et~al.}(2018)Inoue, Shiota, Bamba, \& Park}]{inoue2018magnetohydrodynamic}
Inoue, S., Shiota, D., Bamba, Y., \& Park, S.-H. 2018, The Astrophysical Journal, 867, 83

\bibitem[{{Jiang} {et~al.}(2016){Jiang}, {Wu}, {Feng}, \& {Hu}}]{jiang2016data}
{Jiang}, C., {Wu}, S.~T., {Feng}, X., \& {Hu}, Q. 2016, Nature Communications, 7, 11522, \dodoi{10.1038/ncomms11522}

\bibitem[{{Leake} \& {Arber}(2006)}]{2006A&A...450..805L}
{Leake}, J.~E., \& {Arber}, T.~D. 2006, \aap, 450, 805, \dodoi{10.1051/0004-6361:20054099}

\bibitem[{{Leake} {et~al.}(2017){Leake}, {Linton}, \& {Schuck}}]{leake2017testing}
{Leake}, J.~E., {Linton}, M.~G., \& {Schuck}, P.~W. 2017, \apj, 838, 113, \dodoi{10.3847/1538-4357/aa6578}

\bibitem[{{Leake} {et~al.}(2013){Leake}, {Linton}, \& {T{\"o}r{\"o}k}}]{2013ApJ...778...99L}
{Leake}, J.~E., {Linton}, M.~G., \& {T{\"o}r{\"o}k}, T. 2013, \apj, 778, 99, \dodoi{10.1088/0004-637X/778/2/99}

\bibitem[{{Low}(1991)}]{1991ApJ...370..427L}
{Low}, B.~C. 1991, \apj, 370, 427, \dodoi{10.1086/169829}

\bibitem[{{MacTaggart} \& {Prior}(2021)}]{2021GApFD.115...85M}
{MacTaggart}, D., \& {Prior}, C. 2021, Geophysical and Astrophysical Fluid Dynamics, 115, 85, \dodoi{10.1080/03091929.2020.1740925}

\bibitem[{{MacTaggart} {et~al.}(2021){MacTaggart}, {Prior}, {Raphaldini}, {Romano}, \& {Guglielmino}}]{2021NatCo..12.6621M}
{MacTaggart}, D., {Prior}, C., {Raphaldini}, B., {Romano}, P., \& {Guglielmino}, S.~L. 2021, Nature Communications, 12, 6621, \dodoi{10.1038/s41467-021-26981-7}

\bibitem[{MacTaggart {et~al.}(2021)MacTaggart, Prior, Raphaldini, Romano, \& Guglielmino}]{mactaggart2021direct}
MacTaggart, D., Prior, C., Raphaldini, B., Romano, P., \& Guglielmino, S.~L. 2021, Nature communications, 12, 6621

\bibitem[{{Miyoshi} {et~al.}(2020){Miyoshi}, {Kusano}, \& {Inoue}}]{miyoshi2020magnetohydrodynamic}
{Miyoshi}, T., {Kusano}, K., \& {Inoue}, S. 2020, \apjs, 247, 6, \dodoi{10.3847/1538-4365/ab64f2}

\bibitem[{{Neukirch} \& {Wiegelmann}(2019)}]{2019SoPh..294..171N}
{Neukirch}, T., \& {Wiegelmann}, T. 2019, \solphys, 294, 171, \dodoi{10.1007/s11207-019-1561-0}

\bibitem[{{Pariat} {et~al.}(2009){Pariat}, {Antiochos}, \& {DeVore}}]{2009ApJ...691...61P}
{Pariat}, E., {Antiochos}, S.~K., \& {DeVore}, C.~R. 2009, \apj, 691, 61, \dodoi{10.1088/0004-637X/691/1/61}

\bibitem[{{Price} {et~al.}(2020){Price}, {Pomoell}, \& {Kilpua}}]{price2020exploring}
{Price}, D.~J., {Pomoell}, J., \& {Kilpua}, E.~K.~J. 2020, \aap, 644, A28, \dodoi{10.1051/0004-6361/202038925}

\bibitem[{{Priest} \& {Forbes}(2002)}]{priest2002magnetic}
{Priest}, E.~R., \& {Forbes}, T.~G. 2002, \aapr, 10, 313, \dodoi{10.1007/s001590100013}

\bibitem[{{Rice} \& {Yeates}(2023)}]{2023ApJ...955..114R}
{Rice}, O. E.~K., \& {Yeates}, A.~R. 2023, \apj, 955, 114, \dodoi{10.3847/1538-4357/acefc1}

\bibitem[{{Snodgrass}(1983)}]{1983ApJ...270..288S}
{Snodgrass}, H.~B. 1983, The Astrophysical Journal, 270, 288, \dodoi{10.1086/161121}

\bibitem[{{Syntelis} {et~al.}(2019){Syntelis}, {Lee}, {Fairbairn}, {Archontis}, \& {Hood}}]{syntelis2019eruptions}
{Syntelis}, P., {Lee}, E.~J., {Fairbairn}, C.~W., {Archontis}, V., \& {Hood}, A.~W. 2019, \aap, 630, A134, \dodoi{10.1051/0004-6361/201936246}

\bibitem[{{Temmer}(2021)}]{temmer2021space}
{Temmer}, M. 2021, Living Reviews in Solar Physics, 18, 4, \dodoi{10.1007/s41116-021-00030-3}

\bibitem[{{Toriumi} \& {Hotta}(2019)}]{toriumi2019spontaneous}
{Toriumi}, S., \& {Hotta}, H. 2019, \apjl, 886, L21, \dodoi{10.3847/2041-8213/ab55e7}

\bibitem[{{Toriumi} {et~al.}(2023){Toriumi}, {Hotta}, \& {Kusano}}]{2023NatSR..13.8994T}
{Toriumi}, S., {Hotta}, H., \& {Kusano}, K. 2023, Scientific Reports, 13, 8994, \dodoi{10.1038/s41598-023-36188-z}

\bibitem[{{van Ballegooijen} \& {Martens}(1989)}]{1989ApJ...343..971V}
{van Ballegooijen}, A.~A., \& {Martens}, P.~C.~H. 1989, The Astrophysical Journal, 343, 971, \dodoi{10.1086/167766}

\bibitem[{{van Ballegooijen} {et~al.}(2000){van Ballegooijen}, {Priest}, \& {Mackay}}]{2000ApJ...539..983V}
{van Ballegooijen}, A.~A., {Priest}, E.~R., \& {Mackay}, D.~H. 2000, \apj, 539, 983, \dodoi{10.1086/309265}

\bibitem[{{Vissers} {et~al.}(2022){Vissers}, {Danilovic}, {Zhu}, {Leenaarts}, {D{\'\i}az Baso}, {da Silva Santos}, {de la Cruz Rodr{\'\i}guez}, \& {Wiegelmann}}]{vissers2022active}
{Vissers}, G.~J.~M., {Danilovic}, S., {Zhu}, X., {et~al.} 2022, \aap, 662, A88, \dodoi{10.1051/0004-6361/202142087}

\bibitem[{{Warnecke} \& {Peter}(2019)}]{warnecke2019data}
{Warnecke}, J., \& {Peter}, H. 2019, \aap, 624, L12, \dodoi{10.1051/0004-6361/201935385}

\bibitem[{{Wiegelmann} \& {Madjarska}(2023)}]{2023SoPh..298....3W}
{Wiegelmann}, T., \& {Madjarska}, M.~S. 2023, \solphys, 298, 3, \dodoi{10.1007/s11207-022-02094-2}

\bibitem[{{Wiegelmann} {et~al.}(2015){Wiegelmann}, {Neukirch}, {Nickeler}, {Solanki}, {Mart{\'\i}nez Pillet}, \& {Borrero}}]{2015ApJ...815...10W}
{Wiegelmann}, T., {Neukirch}, T., {Nickeler}, D.~H., {et~al.} 2015, \apj, 815, 10, \dodoi{10.1088/0004-637X/815/1/10}

\bibitem[{{Wiegelmann} \& {Sakurai}(2021)}]{Wiegelmann2021LRSP}
{Wiegelmann}, T., \& {Sakurai}, T. 2021, Living Reviews in Solar Physics, 18, 1, \dodoi{10.1007/s41116-020-00027-4}

\bibitem[{{Yang} {et~al.}(1986){Yang}, {Sturrock}, \& {Antiochos}}]{1986ApJ...309..383Y}
{Yang}, W.~H., {Sturrock}, P.~A., \& {Antiochos}, S.~K. 1986, \apj, 309, 383, \dodoi{10.1086/164610}

\bibitem[{{Yardley} {et~al.}(2018){Yardley}, {Mackay}, \& {Green}}]{yardley2018simulating}
{Yardley}, S.~L., {Mackay}, D.~H., \& {Green}, L.~M. 2018, \apj, 852, 82, \dodoi{10.3847/1538-4357/aa9f20}

\bibitem[{{Yeates}(2014)}]{2014SoPh..289..631Y}
{Yeates}, A.~R. 2014, Solar Physics, 289, 631, \dodoi{10.1007/s11207-013-0301-0}

\bibitem[{{Zhu} {et~al.}(2022){Zhu}, {Neukrich}, \& {Wiegelmann}}]{zhu2022magnetohydrostatic}
{Zhu}, X., {Neukrich}, T., \& {Wiegelmann}, T. 2022, Science in China E: Technological Sciences, 65, 1710, \dodoi{10.1007/s11431-022-2047-8}

\bibitem[{{Zhu} \& {Wiegelmann}(2018)}]{zhu2018extrapolation}
{Zhu}, X., \& {Wiegelmann}, T. 2018, \apj, 866, 130, \dodoi{10.3847/1538-4357/aadf7f}

\end{thebibliography}
\bibliographystyle{aasjournal}



Both authors would like to acknowledge funding from The UK Science and Technology funding council under grant number ST/W00108X/1. The MHD simulations used the LARE2D code from the University of Warwick  (\url{https://github.com/Warwick-Plasma/Lare2d}).

\end{document}